\documentclass[letterpaper]{article}

\usepackage[normalem]{ulem}

\usepackage{aaai}
\usepackage{times}
\usepackage{helvet}
\usepackage{courier}
\usepackage{graphicx}

\usepackage{times}
\usepackage{latexsym}
\usepackage{amsmath}
\usepackage{multirow}
\usepackage{url}
\usepackage{paralist}

\title{Do Linguistic Style and Readability of Scientific Abstracts Affect their Virality?} 

\author{Marco Guerini \\
  Trento-Rise \\
  Via Sommarive 18, Povo \\
  Trento --- Italy \\
 {\tt marco.guerini@trentorise.eu} \\\And
  Alberto Pepe \\
  Center for Astrophysics \\
  Harvard University \\
  Cambridge, MA --- USA \\
  {\tt apepe@cfa.harvard.edu} 
  \\\And
  Bruno Lepri \\
  Media Lab \\
  Massachusetts Institute of Technology  \\
  Cambridge, MA --- USA \\
  {\tt lepri@fbk.eu} \\}

\date{}

\begin{document}
\maketitle
\begin{abstract}
Reactions to textual content posted in an online social network show
different dynamics depending on the linguistic style and readability
of the submitted content. Do similar dynamics exist for responses to
scientific articles? Our intuition, supported by previous research,
suggests that the success of a scientific article depends on its
content, rather than on its linguistic style. In this article, we
examine a corpus of scientific abstracts and three forms of associated
reactions: article downloads, citations, and bookmarks. Through a
class-based psycholinguistic analysis and readability indices tests,
we show that certain stylistic and readability features of abstracts clearly concur in
determining the success and viral capability of a scientific
article. 
\end{abstract}

\section{Introduction}

The  generic term \textit{virality} refers to the tendency of information to spread quickly and widely in a community by word-of-mouth processes. Analyzing and recognizing such forms of persuasive communication is of paramount importance in many theoretical and applied contexts. For example, what determines whether content posted and shared on Facebook, Digg, Google+, or Twitter will go \textit{viral} or not? 

We agree with the view that virality hinges primarily on the nature of
the content being spread
\cite{virality,aral2010creating,marco:carlo:gozde:ICWSM-11}. Yet, when
analyzing text-rich contexts, another important component may
contribute to the viral potential of information: its linguistic
style. Textual snippets posted on social network sites, e.g., status
updates and tweets, may be prone to receive more attention than others
based not only on their content but also on \textit{how} they are
written \cite{querciamood}. Can we assume similar dynamics when analyzing virality of scientific articles? 

\medskip In this paper, we analyze virality in terms of a community's
response to a scientific article measuring the volume of downloads,
bookmarks, and citations it receives. These three indicators are
telling of the extent of penetration of a scientific article in a
given scientific community along three different tangents. Citations
in the scholarly record certainly represent the most widely employed
and accepted measure of validity and visibility in science. Yet, due
to the lengthy time frames of academic publishing, citations are
normally accrued relatively slowly. Readership, as measured in the
total number of clicks or downloads a paper receives is the most
direct and immediate yardstick for visibility. Several readership and
usage measures have been tested and discussed in the literature
\cite[for a general review]{bollen}. Bookmarking is another readily
available indicator of visibility. Websites such as CiteULike
(\url{www.citeulike.org}) allow users to store, organize and share
links to academic papers. These novel measures of impact ---
downloads, social bookmarks, and social media responses --- are
increasingly being adopted by bibliographic services and promise to
play an important role in academic evaluation in the near future \cite{thelwall,pepe}. 

\medskip Text based investigations of scientific virality 
have recently appeared in the literature. Routledge and Smith (\citeyear{routledgepredicting}) analyze corpora of abstracts and fulltexts from  different communities. They consider downloads and within-community citations as response indicators to articles and use generalized linear models to predict them. Their results show that textual features significantly improve accuracy of virality predictions over metadata such as authors, topic, and publication venues. We consider such finding as a starting point for our analysis. In fact, 
rather than focusing on the task of predicting responses, we try to model the non-topical features (i.e. language style and readability) of a viral text, considering \emph{only} the abstract of the paper. As such, we assume that virality is triggered mainly by the abstract of the scientific article. This is a fair assumption considering that the abstract is, by and large, the main vehicle of scientific dissemination and circulation in online digital platforms.

\medskip Our approach can be explained in light of a ``rapid
cognition" model \cite{ambady1992thin,kenny1994interpersonal}. In this
model, the user has to decide in a limited amount of time whether to
download, bookmark, and/or cite a paper. In order to make a decision,
she exploits cues which are not directly related to the content of
the paper such as its readability and writing style, e.g. whether the
text is presented in an assertive way, using self centered
pronouns such as ``we''. In some respects, the rapid cognition model
is reminiscent of the mechanisms by which humans routinely make
judgments about strangers' personality and behavior from very short
behavioral sequences and non-verbal cues. Those intuitions, based on so-called ``thin slices" of behavior, the process they come by, and their effectiveness in producing  
precise judgments on individual's or group's properties (e.g. personality, teaching capabilities, negotiation outcome) have been subject to extensive investigation by social psychologists  \cite{kenny1994interpersonal}.

\subsubsection{Dataset.} Our analysis is based on a corpus of articles in the field of physics and astronomy published in the last decade. The corpus is obtained from the NASA Astrophysics Data System (ADS), a complete database of physics and astronomy literature with a user base which includes virtually every researcher in astrophysics and related disciplines. For each paper in this corpus, we avail of the following information: the text abstract of the paper, the number of times it is downloaded on the ADS website, the number of times it is cited in the literature, the number of times it is bookmarked on the CiteULike website. From this bibliographic corpus we extract three balanced collections of ``viral papers'': (1) the most cited papers (number of cites $\geq$ 350), (2) the most downloaded papers (downloads $\geq$ 330), and (3) the most bookmarked papers (bookmarks $\geq$ 8). An additional collection is also created, containing a random selection of non-viral papers (i.e. papers that scored 0 on the three indicators above), to be used as a ground comparison. Each one of these collections contains roughly 3,000 abstracts. The completeness of the ADS database and its wide adoption rate guarantees that (i) datasets are homogeneous/comparable and (ii) findings about language style, if any, can be tracked back to the viral properties of the abstracts and not to specific communities over-representation in one medium. 

We employ these four datasets to perform two different analyses: (1) a
class-based psycholinguistic analysis and (2) a readability indices
test. The features extracted in the first analysis track back to a
number of psycholinguistic attributes, e.g. the way information is
presented, the use of personal rather than impersonal references to
the work, the use of time-related verb forms, and so on. With the
second analysis, we measure the readability of the abstracts, i.e.,
how difficult it is to understand their language. We demonstrate that there are important features, not directly connected with the content of a paper, which concur in determining its success. 

\section{Class-based psycholinguistic analysis}
To explore the characteristics of viral texts, we employ a class-based psycholinguistic analysis of text which can be adapted to studies of social contagion \cite{rada:carlo:ACL-09}. We calculate a score associated with a given class of words, as a measure of saliency for the given word class inside the collection of most \textit{cited}, \textit{downloaded} and  \textit{bookmarked} articles.

Given a class of words $C=\{W_1,W_2,...,W_N\}$, we define the class coverage in the viral abstract collection $A$ as the percentage of words from $A$ belonging to the class $C$:

{\small
\begin{equation}\tag{COV}\label{eq: coverage}
Coverage_{A}(C) =  \frac{\sum_{W_{i} \in C} Frequency_A(W_i)}{Size_A}
\end{equation}
}
where $Frequency_A(W_i)$ represents the number of occurrences of word $W_i$ inside corpus $A$, and $Size_A$ represents the total size (in words) of the corpus $A$. Similarly, we define class $C$ coverage for the corpus of control abstracts $\mathcal{D}$.

The \textit{dominance score} of the class $C$ in the given corpus $A$ is then defined as the ratio between the coverage of the class in the examples set $A$ with respect to the coverage of the same class in
the corpus $\mathcal{D}$:

{\small
\begin{equation}\tag{DOM}\label{eq: coverage}
Dominance_A(C) = \frac{Coverage_A(C)}{Coverage_\mathcal{D}(C)}
\end{equation}
}

A dominance score higher than 1 indicates a class that is
dominant in collection $A$.  
A score lower than 1 indicates 
a class that is  
unlikely to appear in collection $A$.
We use the classes of words as defined in the Linguistic Inquiry and Word Count (LIWC),
which was developed  
for psycholinguistic analysis \cite{Pennebaker99}.
LIWC includes about 2,200 words and word stems grouped into about 70
broad categories relevant to psychological processes
(e.g., EMOTION, COGNITION). Sample words for relevant classes in our study are shown in Table \ref{categories-table}.

\begin{table}[h!]
{\footnotesize
\begin{center}
\begin{tabular}{l|l} 
\hline\hline
\emph{LABEL}  &  \emph{Sample words}  \\ \hline 
\textbf{CERTAIN} & all, very, fact*, exact*, certain*, completely\\
\textbf{NEGATE} & not, no, zero, without, never\\
\textbf{DISCREP} & but, if, expect*, should\\
\textbf{TENTAT} & or, some, may, possib*, probab*\\
\textbf{SENSES} & observ*, discuss*, shows, appears\\
\textbf{SELF} & we, our, I, us\\
\textbf{SOCIAL} & discuss*, interact*, suggest*, argu*\\
%
\hline\hline
\end{tabular}
\end{center}
}
\caption{\label{categories-table} Word categories along with a sample of corresponding most frequent words in the datasets}
\end{table}

\subsubsection{Results and Discussion}
Tables \ref{words-table} and \ref{dominace-partial-table} show top
ranked classes along with dominance scores. In the following
 we  clustered these classes according to macro-categories
that emerged from the analysis. To keep only significant results, we
made a cutoff for dominance scores included between 1.2 and 0.8, as
proposed by Mihalcea and Strapparava (\citeyear {rada:carlo:ACL-09}).

\begin{table}[htb]

{\footnotesize
\begin{center}
\begin{tabular}{l|lll} 
\hline\hline
& Downl & Bookm & Cite \\ \hline
CERTAIN & 1.58 & 1.51 & 1.65 \\
DISCREP & 1.88 & 1.94 & 1.71 \\
EXCL & 1.51 & 1.80 & 1.26 \\
FUTURE & 1.25 & 1.40 & 1.54 \\
NEGATE & 1.33 & 1.42 & 1.33 \\
OTHREF & 3.06 & 2.77 & 1.62 \\
PAST & 0.40 & 0.64 & 0.53 \\
PRONOUN & 2.19 & 2.09 & 1.48 \\
SELF & 3.56 & 2.93 & 1.82 \\
SENSES & 1.53 & 1.23 & 1.32 \\
SIMILES & 1.30 & 1.21 & 1.54 \\
SOCIAL & 1.94 & 2.64 & 1.63 \\
TENTAT & 1.36 & 1.76 & 1.56 \\
WE & 3.70 & 3.07 & 1.84 \\
\hline\hline
\end{tabular}
\end{center}
}
\caption{\label{words-table} Dominant Word Classes in all three datasets}

\end{table}

Basically, our approach consists in counting words in psychologically meaningful
categories. 
The LIWC was created for spontaneous, personal language production. Since we are analyzing scientific texts (non spontaneous, by
definition), we ruled out those categories that are more focused on
content rather than style (e.g. RELIGION, MUSIC), because their relevance can be connected to the polysemy of the corresponding words rather than a
presence of the category itself in the abstract. As an example,
consider the words "disk", "radio*", "band", "instrument*" from the
LIWC MUSIC category: in the physics and astronomy field these words have a completely different meaning.

\subsubsection{Categories of Basic Virality.}

We begin by analyzing those
categories that are dominant in all the three datasets, from Table \ref{words-table}. These
categories represent a basic
form of virality, common to all datasets.

\emph{Certainty Dimension.} We found a significant dominance of categories describing cognitive
processes (in particular the style of presentation of a given content). Viral papers tend to use, in the abstract, polarized forms
of such way of presenting information. On the one side, a more assertive
language (CERTAIN) is found - also in the negative form (NEGATE). On the
other side, certainty is mitigated by showing discrepancies between
what was expected
and what was actually found (DISCREP), highlighting the boundaries of assertions coverage 
 (EXCL - e.g. but, except, without). 
Interestingly the assertive language is also mitigated by the category
expressing tentative standpoints (TENTAT).

\emph{Time-related Dimension.} With regard to time-related language style we see a positive
correlation with verbs in the future form (FUTURE) and
a negative correlation with verbs in the past form (PAST).

\emph{Self-centered Dimension.} Viral articles are usually presented in a personal rather than
impersonal way, not only in the general use of pronouns (PRONOUN) but specifically 
in the use of self centered pronouns, representing the researchers (SELF, and in particular WE).

\emph{Sense-related and other Dimensions.}  Finally, in viral papers we observe the tendency of describing the work through sense-related rather than abstract verbs (SENSES), using
similitudes (SIMILES, e.g.  like) and using terms related
to social interaction (SOCIAL). 

\subsubsection{Categories of Specialized Virality.}

We also analyze 
those categories that are dominant in only some datasets or that are representative of a specific dataset. Results are summarized in Table \ref{dominace-partial-table}.

\emph{Certainty Dimension.}  Frequently downloaded papers use
less often terms related to achievements (ACHIEVE) and more often 
terms in the ASSENT category (agree*, indeed, accepta*), when compared
to the control dataset. In general the most bookmarked dataset is the only one having a positive correlation with the macro-class of cognitive mechanisms (COGMECH), due to the further correlation with INHIBIT and INSIGHT. 

\emph{Time-related Dimension.} Only most bookmarked articles show a positive correlation with verbs in the present form.

\emph{Self-centered Dimension.} Most downloaded and cited articles tend to use more often also self centered pronouns representing the researcher in the first person (I), while most downloaded and most bookmarked papers tend to compare with other researchers' work (OTHER - their*, they, them). 

\emph{Sense-related and other Dimensions.} We notice that 
the use of sense-related verbs diverges on the specific senses when
considering the single viral phenomena (SEE, HEAR, FEEL). The use of
terms related to social interaction (SOCIAL) is further specialized in
verbs concerning communication (COMM) in the most bookmarked and cited
datasets. 


\begin{table}[hb]
{\footnotesize
\begin{center}
\begin{tabular}{llll} 

\hline\hline
& Downl & Bookm & Cite \\
\hline
ACHIEVE & \textbf{0.79} & 0.98 & 0.91 \\
ASSENT & \textbf{1.58} & \textbf{0.71} & \textbf{1.23} \\
COGMECH & 1.04 & \textbf{1.28} & 1.06 \\
ARTICLE & 0.93 & \textbf{0.77} & 1.03 \\
COMM & 1.04 & \textbf{1.93} & \textbf{1.95} \\
FEEL & \textbf{0.35} & 0.99 & 1.12 \\
HEAR & 0.72 & \textbf{1.20} & \textbf{2.04} \\
SEE & \textbf{1.91} & \textbf{1.25} & 1.15 \\
I & \textbf{1.76} & 1.12 & \textbf{1.65} \\
OTHER & \textbf{1.68} & \textbf{2.09} & 1.16 \\
INHIB & 1.00 & \textbf{1.39} & \textbf{1.21} \\
INSIGHT & 0.97 & \textbf{1.22} & 1.00 \\
PRESENT & 1.04 & \textbf{1.28} & 1.25 \\

\hline\hline
\end{tabular}
\end{center}
}
\caption{\label{dominace-partial-table} Dominant Word Classes in some datasets}
\end{table}

\section{Readability Index Tests}

We further analyzed the abstracts in the three datasets according to readability indices, to understand whether there is a difference in the language difficulty among them. Basically, the task of readability assessment consists in quantifying how difficult a text is for a reader. This kind of assessment has been widely used for several purposes, such as evaluating the reading level of children and impaired persons and improving Web content accessibility. 

We use two indices to compute the difficulty of an abstract: the
Gunning Fog \cite{gunning-52} and the Flesch indices \cite{flesch-46}. These metrics combine factors, such as word and sentence length, that are easy to compute and approximate the linguistic elements that impact on readability. 

The \textbf{Fog index} is a rough measure of how many years of
schooling it would take someone to understand the content; higher
scores indicate material that is harder to read. 
Texts requiring near-universal understanding have an index less than 8. Academic papers usually have a score between 15 and 20. 

The \textbf{Flesch Index} rates  texts on a 100-point scale. Higher
scores indicate material that is easier to read while lower numbers
mark passages that are more difficult to read. Scores can be
interpreted as: 90-100 for content easily understood by an average
11-year-old student, while 0-30 for content best understood by university graduates.

\begin{table}[htb]
{\footnotesize
\begin{center}
\begin{tabular}{lrrrr} 

\hline\hline
& \multicolumn{2}{c}{Fog-index} & \multicolumn{2}{c}{Flesch-index}\\ 
& \multicolumn{1}{c}{$\mu$} & \multicolumn{1}{c}{$\sigma$} & \multicolumn{1}{c}{$\mu$} & \multicolumn{1}{c}{$\sigma$}\\
\hline
Bookm & \textbf{21.02*} & 3.37 & \textbf{8.77*} & 14.44\\ 
Cites & \textbf{19.83$^\dagger$} & 4.03 & \textbf{15.80$^\dagger$} & 15.10\\ 
Downl & \textbf{18.22*} & 3.86 & \textbf{25.86*} & 13.48\\ 
Control & \textbf{19.95} & 4.18 & \textbf{14.80} & 15.96\\ 
\hline\hline
\end{tabular}
\end{center}
}
\caption{\label{readability-indexes-table} Averaged readability indexes for the various datasets. * means a statistically significant difference at $\alpha$ $<$  0.001,  $^\dagger$ means no statistically significant difference, with respect to the control dataset. T-test used.}
\end{table}

\subsubsection{Results and Discussion.}

As expected all abstracts have high-difficulty readability scores (see table \ref{readability-indexes-table}). But interestingly, while most cited papers have scores that are not statistically different from baseline papers, most bookmarked papers have abstracts that are harder to read and most downloaded papers have abstracts easier to read. Furthermore, the standard deviation tend to diminish in most-bookmarked and most-downloaded papers, indicating that these classes tend to converge in readability difficulty (F-test, $\alpha$ $<$  0.001).\medskip 


These results suggest different practices/uses associated with the
various datasets, in line with the assumption that virality is a phenomenon with many facets \cite{marco:carlo:gozde:ICWSM-11}. These practices can be possibly interpreted as steps of a process that goes from initial interest/curiosity for an article to the final decision of citing it: 
(i) The most downloaded papers are those that are easier to read and probably get more initial attention and understanding. 
(ii) On the contrary the most bookmarked are those that need a deeper understanding and so are "put in the stack" to be analyzed later on. 
(iii) Finally, being cited is much less connected to readability (indicating that what matters in the end is the style and content of the abstract/paper). 


\section{Conclusions}

In this paper we argued that responses to scientific articles are influenced by the linguistic style and readability of their abstracts. Through a psycholinguistic analysis and readability tests, we showed that linguistic style of abstracts concurs in determining the success 
of a scientific article. Based on these findings, we modified the
initial abstract of the present paper, so to meet virality criteria of Table 2 (key modifications are underlined, added text in bold): 

\begin{quotation}
{\footnotesize \noindent Reactions to textual content posted in an online
  social network show different dynamics \sout{hinging}
  \underline{depending} on the linguistic style and readability of the submitted content. Do similar dynamics exist for responses to scientific articles? \sout{The}  \underline{Our} intuition, \underline{supported} \textbf{by previous research}, \sout{says} \underline{suggests} that the success of a scientific article 
depends on its content, \underline{rather} \textbf{than on its
  linguistic style}. In this article, \underline{we} \textbf{examine}
a corpus of scientific abstracts and three forms of associated
reactions \sout{is examined}: article downloads, citations, and
bookmarks. Through a class-based psycholinguistic analysis and
readability indices tests, we \sout{argue} \underline{show} that
certain stylistic and readability features of abstracts \underline{clearly} concur in
determining the success and viral capability of a scientific
article.}
\end{quotation}

The final version of the abstract showed a significant dominance  on
72\% of the Word Classes presented in Table 2 (57\% before the
modification) and its readability scores (unchanged) are 18.81 (Fog-index) and 22.57 (Flesch-index).

\section{Acknowledgment}
We would like to thank Alberto Accomazzi, Jay Luker and the
ADS Team for providing access to the
bibliographic, citation, and usage data. We
thank Sara Tonelli for advising on the readability indices analysis. 
Marco Guerini was partially supported by a Google Research Award. 
Bruno Lepri's research is funded by PERSI project inside the Marie Curie COFUND 7th Framework.

\bibliographystyle{aaai} 
\bibliography{Persuasive} 

\end{document}